\documentclass{ifacconf}

\usepackage{graphicx}      %

\makeatletter
\let\old@ssect\@ssect %
\makeatother

\usepackage{natbib}        %

\usepackage{amsmath,amssymb} 

\usepackage{color}
\usepackage[usenames]{xcolor}
\usepackage{tikz}
\tikzset{beameritem/.style={circle,inner sep=0,minimum size=2ex,text=enumerate item.bg,fill=enumerate item.fg,font=\footnotesize}}
\usepackage{pgfplots} %
\pgfplotsset{compat=newest}
\usepackage{pgfpages}
\usepackage{siunitx}

\usepackage{hyperref}
\makeatletter
\def\@ssect#1#2#3#4#5#6{%
	\NR@gettitle{#6}%
	\old@ssect{#1}{#2}{#3}{#4}{#5}{#6}%
}
\makeatother

\hypersetup{
	colorlinks,
	linkcolor={black},
	citecolor={black},
	urlcolor={black}
}

\usetikzlibrary{backgrounds}
\usetikzlibrary{shapes}

\newcommand{\mati}{\tau_{\text{MATI}} }
\newcommand{\defeq}{:=}

\newcommand{\change}[2]{#2}
\newcommand{\changet}[2]{{#2}}
\newcommand{\percent}{\%}

\newcommand{\norm}[1]{\left\lVert#1\right\rVert}
\newcommand{\abs}[1]{\left|#1\right|}

\newtheorem{sasum}{Standing Assumption}
\newtheorem{defi}{Definition}
\newtheorem{asum}{Assumption}
\newtheorem{rema}{Remark}
\newtheorem{theo}{Theorem}
\newtheorem{coro}{Corollary}
\begin{document}
	\begin{frontmatter}
		
		\title{A Simple Approach to Increase the Maximum Allowable Transmission Interval\thanksref{footnoteinfo}} 

		\thanks[footnoteinfo]{\changet{The authors thank the German Research Foundation (DFG) for support of this work within grant AL 316/13-2  and within the German Excellence Strategy under grant EXC-2075 -
			285825138; 390740016.}{
			$\copyright$ 2021 the authors. This work has been accepted to IFAC for publication under a Creative Commons Licence CC-BY-NC-ND.\\
			 Funded by Deutsche Forschungsgemeinschaft (DFG, German Research Foundation) under Germany’s Excellence Strategy - EXC 2075 - 390740016 and under grant AL 316/13-2 - 285825138. We acknowledge the support by the Stuttgart Center for Simulation Science (SimTech).}}
		
		\author[first]{Michael Hertneck} 
		\author[first]{Frank Allg\"ower} 
		
		\address[first]{University of Stuttgart, Institute for Systems Theory and Automatic Control, 70569 Stuttgart, Germany (email: $\{$hertneck,allgower$\}$@ist.uni-stuttgart.de).}
		
		\begin{abstract}                %
			When designing Networked Control Systems (NCS), the maximum allowable transmission interval (MATI) is an important quantity, as it provides the admissible time between two transmission instants. An efficient procedure to compute a bound on the MATI  such that stability can be guaranteed for general nonlinear NCS is the emulation of a continuous-time controller. In this paper, we present a simple but efficient modification to the well-established emulation-based approach from \cite{carnevale2007lyapunov} to derive a bound on the MATI. Whilst only minor \change{}{technical } changes are required, the proposed modification can lead to significant improvements for the MATI bound as compared to \cite{carnevale2007lyapunov}. We revisit two numerical examples from literature and demonstrate that the improvement may amount to more than 100\percent. 
		\end{abstract}
		
		
	\end{frontmatter}	

\section{Introduction}
\label{sec_intro}
Networked Control Systems (NCS) are dynamical systems, where hard-wired links in the feedback loop are replaced by a shared communication medium. A networked architecture for control systems has several advantages, as, e.g., lower cost of installation and maintenance and a greater flexibility. In turn, network-induced effects like a limited packet rate and varying transmission intervals need to be considered when designing NCS (\cite{hespanha2007survey}). An important research topic in the field of NCS is therefore to characterize conditions on the communication channel and on the transmission protocol of the network such that stability guarantees can be obtained for the NCS despite network effects. When considering time-varying sampling intervals, stability guarantees can typically be given as long as the time between two transmissions stays below a maximum allowable transmission interval (MATI). Being able to guarantee stability for a preferably large MATI allows to save network resources and is therefore important in the design of NCS.
Several approaches, based on a wide variety of methods have been developed to derive such bounds on the MATI, 
 cf. \cite{hetel2017recent} for an overview. An efficient procedure is to emulate a continuous-time controller, i.e., to design first the controller assuming perfect communication and to derive then in a second step a bound on the MATI, taking into account the communication channel and the transmission protocol (\cite{walsh2002stability}). For general nonlinear NCS, an emulation-based approach for computing a bound on the MATI, such that stability can be guaranteed, has been proposed in \cite{nesic2004input} and improved in \cite{carnevale2007lyapunov,nesic2009explicit} using hybrid Lyapunov functions. The wide variety of work that is based on the findings from  \cite{carnevale2007lyapunov,nesic2009explicit} illustrates the importance of the approach, see  \cite{heemels2010networked,postoyan2014tracking,dolk2017output} for some examples. There have also been various attempts to improve the results from \cite{carnevale2007lyapunov,nesic2009explicit}. Most approaches are restricted to special system classes or require additional assumptions for the communication network, see, e.g., \cite{bauer2012stability,hertneck20stability,heijmans2020average}. Recently, a notable improvement for the MATI bound of up to 15~\percent~for the setup from \cite{carnevale2007lyapunov} could be achieved in \cite{heijmans2017computing,heijmans2018generalized}, using more general  hybrid Lyapunov function approaches.

In this paper, we \change{present}{propose } a simple modification of the \change{setup}{main result } from \cite{carnevale2007lyapunov} that leads to a significantly improved bound on the MATI for which stability is guaranteed. 
\change{To achieve the improvement, we slightly generalize the main assumption from \cite{carnevale2007lyapunov}. Since the modification is only minor, the refinements required for the approach from \cite{carnevale2007lyapunov} and the respective proofs are straight forward. Still, the MATI bound can be increased significantly, using the modified approach for some examples even by more than 100\percent. Even though the modifications are much simpler than those from \cite{heijmans2017computing,heijmans2018generalized}, the obtained improvements are significantly higher. Moreover, the modification can equivalently be applied to the setup and the results from \cite{nesic2009explicit} and is thus also suited for sampled-data systems. We illustrate the proposed modification with two numerical examples from \cite{nesic2009explicit,heijmans2017computing}, and demonstrate that it leads to an improvement of more that 100\percent~
	in comparison to the approach from \cite{carnevale2007lyapunov,nesic2009explicit} in some situations.
	In addition to the approach for general nonlinear NCS, we demonstrate how our main assumption can be stated as a linear matrix inequality (LMI) for the case of linear NCS, leading to a computationally tractable condition.}{In particular, we relax \cite[Assumption~1]{carnevale2007lyapunov} by considering a more general function $H(x,e)$ instead of restricting ourselves to $H(x)$. Using the proposed modification, the MATI bound can be increased significantly, for some typical benchmark examples from literature even by more than 100\percent. Hence, the proposed modification is very beneficial to determine a large bound on the MATI. In the proofs of \cite{carnevale2007lyapunov}, only minor modifications are required due to the more general choice of $H(x,e)$. This fact is also of great importance, as it makes it easy to apply the modification to many results that build on the approach from \cite{carnevale2007lyapunov}, as, e.g., \cite{nesic2009explicit,heemels2010networked,postoyan2014tracking,dolk2017output}, leading there also to significant improvements.

We illustrate the proposed modification with \changet{three}{two} numerical benchmark examples from literature, and demonstrate for this examples that it leads to an improvement in the range of 66-108\percent~in comparison to the approach from \cite{carnevale2007lyapunov,nesic2009explicit}.
Additionally, we demonstrate for linear systems how our approach can be used systematically to improve the MATI in comparison to \cite{carnevale2007lyapunov,nesic2009explicit}.
	}

The remainder of this paper is structured as follows. First, we recap the setup and problem formulation from \cite{carnevale2007lyapunov} 
\change{in the remainder of this Section and }{}
in Section~\ref{sec_setup}. Then, we present our approach to compute the MATI, give stability guarantees and illustrate the approach with the nonlinear example from \cite{nesic2009explicit} in Section~\ref{sec_main}. In Section~\ref{sec_lin}, we refine our main result for linear NCS using an LMI, and revisit the example from \cite{heijmans2017computing}. The paper is concluded in Section~\ref{sec_conc}.

\subsubsection*{Notation and Definitions}
The positive real numbers are denoted by $\mathbb{R}_{>0}$ and  $\mathbb{R}_{\geq 0} \defeq \mathbb{R}_{> 0} \cup \{0\} $. The positive  natural numbers are denoted by $\mathbb{N}$, and $\mathbb{N}_0\defeq\mathbb{N}\cup  \left\lbrace 0 \right\rbrace $. A continuous function $\alpha: \mathbb{R}_{\geq 0} \rightarrow \mathbb{R}_{\geq 0}$ is a class $ \mathcal{K}$ function if it is strictly increasing and $\alpha(0) = 0$. It is a class $\mathcal{K}_\infty$ function if it is of class $\mathcal{K}$ and it is unbounded. A continuous function $\beta:\mathbb{R}_{\geq 0}\times \mathbb{R}_{\geq 0} \rightarrow \mathbb{R}_{\geq 0}$ is a class $\mathcal{K}\mathcal{L}$ function, if $\beta(\cdot,t)$ is of class $\mathcal{K}$ for each $t \geq 0$ and $\beta(s,\cdot)$ is nonincreasing and satisfies $\lim\limits_{t \rightarrow \infty} \beta(s,t) = 0$ for each $s \geq 0$. A function $\beta:\mathbb{R}_{\geq 0}\times \mathbb{R}_{\geq 0} \times \mathbb{R}_{\geq 0} \rightarrow \mathbb{R}_{\geq 0}$ is a class $\mathcal{K}\mathcal{L}\mathcal{L}$ function if for each $r \geq 0$, $\beta(\cdot,r,\cdot)$ and $\beta(\cdot,\cdot,r)$ belong to class $\mathcal{K}\mathcal{L}$. We denote by $\norm{\cdot}$ the Euclidean norm.

	As in \cite{carnevale2007lyapunov,nesic2009explicit}, we will use the following definitions, that are originally taken from \cite{goebel2006solutions}, to characterize a hybrid model of the considered NCS.
\begin{defi}
	\cite{carnevale2007lyapunov} A \textit{compact hybrid time domain} is a set $\mathcal{D} \subset \mathbb{R}_{\geq0} \times \mathbb{N}_{0}$ given by:
	\begin{equation*}
		\mathcal{D} = \bigcup^{J-1}_{j = 0} \left(\left[t_j,t_{j+1}\right],j\right)
	\end{equation*}
	where $J \in \mathbb{N}_{0}$ and $0 = t_0 \leq t_1 \dots \leq t_J$. A \textit{hybrid time domain} is a set $\mathcal{D} \subset \mathbb{R}_{\geq0} \times \mathbb{N}_{ 0}$ such that, for each $\left(T,J\right) \in \mathcal{D}, \mathcal{D} \cap \left(\left[0,T\right] \times \left\lbrace 0,\dots,J\right\rbrace\right)$ is a compact hybrid time domain.
\end{defi}
\begin{defi}
	\cite{carnevale2007lyapunov} A \textit{hybrid trajectory} is a pair (dom~$\xi$, $\xi$) consisting of the hybrid time domain dom~$\xi$ and a function $\xi$ defined on dom~$\xi$ that is continuously differentiable in $t$ on (dom~$\xi$)$\cap\left(\mathbb{R}_{\geq 0} \times \left\lbrace j \right\rbrace \right)$ for each $j\in\mathbb{N}_{ 0}$. 
\end{defi}
\begin{defi}
\cite{carnevale2007lyapunov}	For the hybrid system $\mathcal{H}$ given by the open state space $O \subset \mathbb{R}^n$ and the data $\left(F,G,C,D\right)$ where $F:O\rightarrow \mathbb{R}^n$ is continuous, $G:O\rightarrow O$ is locally bounded and $C$ and $D$ are subsets of $O$, a hybrid trajectory $\xi:$ dom~$\xi \rightarrow O$ is a solution to $\mathcal{H}$ if
\begin{enumerate}
	\item for all $j \in \mathbb{N}_{ 0}$ and for almost all $t\in I_j \defeq$ dom~$\xi\cap \left(\mathbb{R}_{\geq 0} \times \left\lbrace j \right\rbrace\right),$ we have $\xi(t,j) \in C$ and $\dot{\xi}(t,j) = F(\xi(t,j))$;
	\item for all $(t,j)\in$ dom~$\xi$ such that $(t,j+1) \in $ dom~$\xi$, we have $\xi(t,j)\in D$ and $\xi(t,j+1) = G(\xi(t,j))$.
\end{enumerate}
\end{defi}
For further details on these definitions, see \cite{goebel2006solutions}. Omitting time arguments, the hybrid system model is described by
\begin{equation}
\label{eq_hybrid}
\mathcal{H} = \begin{cases}
\dot{\xi} = F(\xi) & \xi \in C\\
\xi^+ = G(\xi) & \xi \in D,
\end{cases}
\end{equation}
where $\xi^+$ denotes  $\xi(t_{j+1},j+1)$. 
Note that typically $C\cap D \neq \emptyset$ and therefore, the hybrid model we consider may have nonunique solutions.
\section{Setup}
\label{sec_setup}
We consider the same NCS model as in \cite{carnevale2007lyapunov}. The dynamics of plant and controller are given by
\begin{equation}
	\begin{split}
	\dot{x}_P &= f_P(x_P,\hat{u})\\
	y &= g_P(x_P)
	\end{split}
	\label{eq_plant}
\end{equation}
and
\begin{equation}
\begin{split}
\dot{x}_C &= f_C(x_C,\hat{y})\\
u &= g_C(x_C)
\end{split}
\label{eq_controller}
\end{equation}
where $x_P\in\mathbb{R}^{n_P}$ and $x_C \in \mathbb{R}^{n_C}$ denote the plant and controller states, $y \in \mathbb{R}^{n_y}$ is the plant output, $u \in \mathbb{R}^{n_u}$ is the controller output and $\hat{y} \in \mathbb{R}^{n_y}$ and $\hat{u} \in \mathbb{R}^{n_u}$ are the most recent values of $y$ and $u$ that have been received by the controller or respectively by the actuator. We define the network-induced error as 
\begin{equation*}
e(t) \defeq \begin{pmatrix}
\hat{y}(t)-y(t)\\
\hat{u}(t)-u(t)
\end{pmatrix} = \begin{pmatrix}
e_y\\e_u
\end{pmatrix}.
\end{equation*}

 The sensors and actuators of the plant are spread over $l$ nodes with index $i\in\left\lbrace 1,2,\dots,l \right\rbrace$. Let the sequence of transmission times be given by $(t_j), ~j \in \mathbb{N}_{0}$, satisfying $ \epsilon \leq t_{j+1} - t_j \leq \tau_{\max}$ for all $j \in \mathbb{N}_0$ and some fixed $0 < \epsilon \leq \tau_{\max}$. Here, 
 $\tau_{\max}$ is the \textit{maximum allowable transmission interval} (MATI) and 
 $\epsilon$ is an arbitrarily small bound that excludes Zeno behavior.  At each transmission time $t_j$, a scheduling protocol grants network access to one of the nodes. Depending on which node was granted network access, $\hat{y}$ and $\hat{u}$ are updated at transmission times as 
\begin{equation*}
	\begin{split}
	\hat{y}(t_j^+) &= y(t_j) + h_y(i,e(t_j))\\
	\hat{u}(t_j^+) &= u(t_j) + h_u(i,e(t_j)),
	\end{split}
\end{equation*}
where $i$ is the index of the respective node.
The function $h \defeq \left[h_y^\top, h_u^\top\right]^\top$ with $h:\mathbb{N}_0 \times \mathbb{R}^{n_y+n_u} \rightarrow  \mathbb{R}^{n_y+n_u}$ thus models the scheduling protocol \cite{walsh2002stability,nesic2004input}. For many protocols as, e.g., try-once-discard (TOD) and Round Robin (RR), the part of the state error that is measured by the node with network access is reset to $0$.

Combining $x\defeq \left[x_P^\top, x_C^\top\right]^\top$ and $e\defeq \left[e_y^\top, e_u^\top\right]^\top$, we can write the overall NCS in the (even more general) form
\begin{align}
	\dot{x} &= f(x,e)\quad\forall t \in \left[t_{j-1},t_j\right] \label{eq_NCS1} \\
	\dot{e} &= g(x,e)\quad\forall t \in \left[t_{j-1},t_j\right] \label{eq_NCS2} \\
	e\left(t_j^+\right) &=h(j,e(t_j)),\label{eq_NCS3}
\end{align}
for all $j\in\mathbb{N}_0$, where $x \in \mathbb{R}^{n_x }$ and $e\in\mathbb{R}^{n_e}$ with $n_x = n_P+n_C$ and $n_e = n_y+n_u$. If $\hat{y}$ and $\hat{u}$ are kept constant between transmission times, i.e., for the zero-order-hold (ZOH) case with
\begin{equation*}
\begin{split}
\dot{\hat{y}} &=0\quad t \in \left[t_{j-1},t_j\right]\\
\dot{\hat{u}} &=0\quad t \in \left[t_{j-1},t_j\right]
\end{split}
\end{equation*}  for all 
 $j$,  we obtain 
\begin{equation}
\label{eq_def_fg}
\begin{split}
	f(x,e) &= \begin{bmatrix}
	f_P(x_P,g_C(x_C)+e_u)\\
	f_C(x_C,g_P(x_P)+e_y)
	\end{bmatrix}\\
	g(x,e) &= \begin{bmatrix}
	-\frac{\partial g_P}{\partial x_P}f_P(x_P,g_C(x_C)+e_u)\\
	-\frac{\partial g_C}{\partial x_C}f_C(x_C,g_P(x_P)+e_y)
	\end{bmatrix}.\\
\end{split}
\end{equation}
The problem that we consider is the same as in \cite{carnevale2007lyapunov}. 
\begin{prob}
	\label{prob_1}
	\cite{carnevale2007lyapunov} Suppose that the controller \eqref{eq_controller} was designed for the plant \eqref{eq_plant} so that the closed-loop system \eqref{eq_plant}, \eqref{eq_controller} without network is globally asymptotically stable. Determine the value of $\mati$ so that for any $\epsilon \in \left( 0, \mati \right]$ and all $\tau_{\max} \in \left[\epsilon, \mati\right]$, we have that the NCS described by  \eqref{eq_NCS1}, \eqref{eq_NCS2}, \eqref{eq_NCS3} is stable in an appropriate sense. 
\end{prob}

To solve Problem~\ref{prob_1}, we propose an approach similar to the one described in \cite{carnevale2007lyapunov,nesic2009explicit}. The only difference is that we relax \cite[Assumption~1]{carnevale2007lyapunov} slightly. Due to this modification, only minor changes in the proofs of the results from \cite{carnevale2007lyapunov} are required in order to obtain the same stability guarantees as in \cite{carnevale2007lyapunov}. However, this slight modification can lead to improvements of $\mati$ of more than $100\percent$, as we will demonstrate with two numerical examples from \cite{nesic2009explicit} and \cite{heijmans2017computing}. 
The details of our approach are given in the next section.

\section{Main Result}
\label{sec_main}
In this section, we present how the approaches from \cite{carnevale2007lyapunov,nesic2009explicit} can be modified to improve the bound on $\mati$ significantly, and illustrate the improvement with the numerical example from \cite{nesic2009explicit}. The closed-loop NCS~\eqref{eq_NCS1}-\eqref{eq_NCS3} can be written in the form~\eqref{eq_hybrid} as follows. As in \cite{carnevale2007lyapunov,nesic2004input}, we introduce the timer variable $\tau \in\mathbb{R}_{\geq 0}$, which keeps track of the elapsed time since the last transmission and the counter $\kappa \in \mathbb{N}_0$ for the number of transmissions. The resulting system is of the form
\begin{equation}
\label{eq_hybrid_sys}
		\left. \begin{split}
		\dot{x} &= f(x,e)\\
		\dot{e} &= g(x,e)\\
		\dot{\tau} &= 1\\
		\dot{\kappa} &= 0
		\end{split} \right\rbrace \tau \in \left[0,\mati \right],\quad
		\left. \begin{split}
		{x}^+ &= x\\
		{e}^+ &= h(\kappa,e)\\
		{\tau}^+ &= 0\\
		{\kappa}^+ &= \kappa+1
		\end{split} \right\rbrace \tau \in \left[\epsilon,\infty\right].	
\end{equation}
Details on~\eqref{eq_hybrid_sys} can be found in \cite{carnevale2007lyapunov}, where the same model is used. We shall also use Standing Assumption~1 from \cite{carnevale2007lyapunov}.
\begin{sasum}
	\cite{carnevale2007lyapunov} $f$ and $g$ are continuous and $h$ is locally bounded.
\end{sasum}
Note that $\tau$ and $\kappa$ are artificially introduced states. Like in \cite{carnevale2007lyapunov}, we aim therefore at guaranteeing uniform global asymptotic stability for the set $\left\lbrace \left(x,e,\tau,\kappa\right): x = 0, e= 0 \right\rbrace$ as defined next.
\begin{defi}
	\label{def_stab}
	\cite[\change{}{Definition~4}]{carnevale2007lyapunov} For the hybrid system~\eqref{eq_hybrid_sys}, the set $\left\lbrace \left(x,e,\tau,\kappa\right): x = 0, e= 0 \right\rbrace$ is \textit{uniformly globally asymptotically stable} \changet{}{(UGAS)} if there exists $\beta \in \mathcal{K}\mathcal{L}\mathcal{L}$ such that, for each initial condition $\tau(0,0) \in\mathbb{R}_{>0}, \kappa(0,0) \in \mathbb{N}_0, x(0,0) \in \mathbb{R}^{n_x}$, $e(0,0) \in \mathbb{R}^{n_e}$, and each corresponding solution
	\begin{equation}
		\norm{\begin{bmatrix}
			x(t,j)\\
			e(t,j)
			\end{bmatrix}} \leq \beta\left(\norm{\begin{bmatrix}
			x(0,0)\\e(0,0)
			\end{bmatrix}},t,\epsilon j\right)
	\end{equation}
	for all (t,j) in the solutions domain. The set is \textit{uniformly globally exponentially stable} \changet{}{(UGES)}, if $\beta$ can be taken to have the form $\beta(s,t,k) = Ms\exp(-\lambda(t+k))$ for some $M>0$ and $\lambda>0$.
\end{defi} 
\subsection{Improving the MATI}
We shall use the following relaxed version of \cite[Assumption~1]{carnevale2007lyapunov} in order to derive guarantees for \changet{ uniform asymptotic stability}{UGAS} or \changet{uniform exponential stability}{UGES}.
\begin{asum}
	\label{ass_hybrid}
	There exists a function $W : \mathbb{N}_0 \times \mathbb{R}^{n_e} \rightarrow \mathbb{R}_{\geq 0}$ that is locally Lipschitz in its second argument, a locally Lipschitz, positive definite, radially unbounded function $V:\mathbb{R}^{n_x} \rightarrow \mathbb{R}_{\geq 0}$, a continuous function $H: \mathbb{R}^{n_x} \times \mathbb{R}^{n_e} \rightarrow \mathbb{R}_{\geq 0}$,	real numbers
	 $\lambda \in \left(0,1\right), L > 0, \gamma > 0, \underline{\alpha}_W, \overline {\alpha}_W \in \mathcal{K}_{\infty}$ 
	and a continuous, positive definite function $\varrho$ such that, $\forall \kappa \in \mathbb{N}_0$ and  $e \in \mathbb{R}^{n_e}$
	\begin{align}
		\label{eq_w_bound} \underline \alpha_W\left(\norm{e}\right) \leq& W(\kappa,e) \leq \overline\alpha_W\left(\norm{e}\right)\\
		\label{eq_w_jump} W(\kappa+1,h(\kappa,e)) \leq& \lambda W(\kappa,e)
	\end{align}
	and for all $\kappa \in \mathbb{N}_0, x \in \mathbb{R}^{n_x}$ and almost all $e \in \mathbb{R}^{n_e}$
	\begin{equation}
		\left\langle \frac{\partial W(\kappa,e)}{\partial e}, g(x,e) \right\rangle \leq L W(\kappa, e) + H(x,e) \label{eq_h}
	\end{equation}
	moreover, for all $e \in \mathbb{R}^{n_e}$, all $\kappa\in\mathbb{N}_0$ and almost all $x \in \mathbb{R}^{n_x}$,
	\begin{align}
		\nonumber \left\langle \nabla V(x),f(x,e) \right\rangle \leq& -\varrho(\norm{x}) - \varrho(W(\kappa,e)) \\&- H^2(x,e) + \gamma^2 W^2(\kappa,e). \label{eq_V_bound}
	\end{align}
\end{asum}
\change{The difference to \cite[Assumption~1]{carnevale2007lyapunov} is that we allow for a more general function $H(x,e)$ instead of restricting ourselves to $H(x)$. Note that \eqref{eq_V_bound} still establishes an $\mathcal{L}_2$ gain $\gamma$ from $W$ to $H$, however the output $H$ is chosen more general for our approach.}{}
Based on Assumption~\ref{ass_hybrid}, we will show that uniform \changet{global asymptotic stability}{UGAS} can be guaranteed if
\begin{equation}
\label{eq_bound}
\mati \leq \begin{cases}
\frac{1}{Lr} \arctan \left(\frac{r(1-\lambda)}{2 \frac{\lambda}{1+\lambda} \left(\frac{\gamma}{L}-1\right)+1+\lambda}\right) & \gamma > L\\
\frac{1}{L} \frac{1-\lambda}{1+\lambda} & \gamma = L\\
\frac{1}{Lr} \text{\normalfont arctanh} \left(\frac{r(1-\lambda)}{2 \frac{\lambda}{1+\lambda} \left(\frac{\gamma}{L}-1\right)+1+\lambda}\right) & \gamma < L,
\end{cases}
\end{equation}
where 
\begin{equation}
\label{eq_r}
r \defeq \sqrt{\abs{\left(\frac{\gamma}{L}\right)^2 -1}}.
\end{equation}
\change{}{\begin{rema}
	 In Assumption~\ref{ass_hybrid}, \eqref{eq_V_bound} establishes an $\mathcal{L}_2$ gain bound $\gamma$ for the system $\dot{x} = f(x,e)$ from the input $W$ to the output $H$ (cf. \cite[Remark~2]{carnevale2007lyapunov}). Moreover, \eqref{eq_w_bound} establishes a bound $L$ on how fast $W$ grows.  By allowing for a more general function $H(x,e)$ in Assumption~\ref{ass_hybrid} instead of restricting ourselves to $H(x)$ as in \cite{carnevale2007lyapunov,nesic2009explicit}, we consider in Assumption~\ref{ass_hybrid} thus a different $\mathcal{L}_2$ gain. Typically, for the output $H(x,e)$, a smaller $\mathcal{L}_2$ gain can be achieved  than the best $\mathcal{L}_2$ gain for the output $H(x)$, whilst simultaneously the growth bound $L$ for $W$ can be reduced. The more general output for $H(x,e)$ can thus lead to a significantly increased bound on $\mati$.
\end{rema}} 
\change{Note that \eqref{eq_bound} and \eqref{eq_r} are identical to (2) and (3) from \cite{carnevale2007lyapunov}. The improvement of the MATI bound originates therefore from the relaxation of $H$ in Assumption~\ref{ass_hybrid}.
	Whilst at the first glance, this seems to be only a small modification, and requires only minor changes in the proofs of the main results from \cite{carnevale2007lyapunov}, it can have a significant effect on the bound on $\mati$. It typically allows us to choose $L$ and $\gamma$ smaller in comparison to
	\cite{carnevale2007lyapunov}, as we will show later for three benchmark examples.  Since the bound on $\mati$ in \eqref{eq_bound} for given $L$ and $\gamma$ is the same as the one in \cite{carnevale2007lyapunov}, being able to chose $L$ and $\gamma$ smaller means being able to increase the MATI bound. Using our Assumption~\ref{ass_hybrid} instead of \cite[Assumption~1]{carnevale2007lyapunov}, we can  in fact improve the MATI in some cases by more than $100\percent$~in comparison to the value obtained in \cite{carnevale2007lyapunov}, as we will demonstrate later. }{\begin{rema}
		Note that \eqref{eq_bound} and \eqref{eq_r} are identical to (2) and (3) from \cite{carnevale2007lyapunov}. However, the more general function $H(x,e)$ in Assumption~\ref{ass_hybrid} can lead to significantly smaller values for $\gamma$ and $L$ in Assumption~\ref{ass_hybrid} in comparison to the minimum possible values for \cite[Assumption~1]{carnevale2007lyapunov}. Since the value of the bound on $\mati$ in \eqref{eq_bound} gets larger when $\gamma$ and $L$ get smaller (cf. \cite[Remark~6]{nesic2009explicit}), the more general choice of $H(x,e)$ leads then to a significant increase of the bound $\mati$.  We will demonstrate this later for two benchmark examples.
	\end{rema}}
\begin{rema}
	The setup of \cite{wang2019periodic} admits to use a more general function $H(x,e)$. However, this additional degree of freedom is not exploited therein.
\end{rema}
We can now state the equivalent to \cite[Theorem 1]{carnevale2007lyapunov} for our setup.
\begin{theo}
	\label{theo_stab}
	Under Assumption~\ref{ass_hybrid}, if $\mati$ in \eqref{eq_hybrid_sys} satisfies the bound \eqref{eq_bound} and $0\change{\leq}{<} \epsilon \leq \mati$ then, for the system \eqref{eq_hybrid_sys}, the set $\left\lbrace \left(x,e,\tau,\kappa\right): x = 0, e= 0 \right\rbrace$ is \changet{uniformly globally asymptotically stable}{UGAS}. If, in addition, there exist strictly positive real numbers $\underline{\alpha}_W, \overline{\alpha}_W, a_1, a_2$ and $a_3$, such that $\underline{\alpha}_W \norm{e} \leq W(\kappa,e) \leq \overline{\alpha}_W \norm{e}$, $a_1\norm{x}^2 \leq V(x) \leq a_2 \norm{x}^2,$ and $\varrho(s) \leq a_3 s^2$, then this set is \changet{uniformly exponentially stable}{UGES}. 
\end{theo}
\begin{pf}
	See Appendix~\ref{append_a}. \hfill\hfill\qed
\end{pf} 
Note that the wording of Theorem~\ref{theo_stab} is identical to the wording of  \cite[Theorem~1]{carnevale2007lyapunov}. \change{the difference is hence only the definition of $H$ in Assumption~\ref{ass_hybrid}. }{A difference to \cite[Theorem~1]{carnevale2007lyapunov} with significant impact on the bound $\mati$ is however the choice of $H$ in Assumption~\ref{ass_hybrid}.}
\change{}{\begin{rema}
	The fact that only minor changes are required in the proof of Theorem~\ref{theo_stab} in comparison to \cite[Theorem~1]{carnevale2007lyapunov} implies that the same modification can also be used for many results that build on \cite{carnevale2007lyapunov}, as e.g. \cite{heemels2010networked,postoyan2014tracking,dolk2017output}.
\end{rema}}
\begin{rema}
	To obtain a sufficient condition for uniform \textit{local} asymptotic stability based on Assumption~\ref{ass_hybrid},  Theorem~2 from \cite{carnevale2007lyapunov} and its proof can easily be modified using Assumption~\ref{ass_hybrid} instead of \cite[Assumption~1]{carnevale2007lyapunov}. 
\end{rema}
\begin{rema}
	For sampled-data systems, i.e., for systems for which $e(t_j^+) = 0$, which implies that \eqref{eq_w_jump} holds even with $\lambda = 0$, the stability results from Theorem~\ref{theo_stab} apply for any $\mati$ that satisfies
	\begin{equation}
		\label{eq_bound_sd}
		\mati < \begin{cases}
		\frac{1}{Lr} \arctan \left(r\right) & \gamma > L\\
		\frac{1}{L}  & \gamma = L\\
		\frac{1}{Lr} \text{\normalfont arctanh} \left(r\right) & \gamma < L
		\end{cases}
	\end{equation}
	with $r$ according to \eqref{eq_r}. The modifications, that are required for this case in our setup and in the proof of Theorem~\ref{theo_stab}, are similar as those in \cite{nesic2009explicit}.
\end{rema}
We are now ready \change{to substantiate our initial assertion, which was}{to demonstrate for the example from \cite{nesic2009explicit} } that the \change{presented}{proposed } modification \change{}{for $H$ } can lead to a significantly larger bound on $\mati$ in comparison to the approach from \cite{carnevale2007lyapunov,nesic2009explicit}.
\change{, using the example from \cite{nesic2009explicit}.}{}
\subsection{Example\change{}{~1}}
We employ the example system from \cite{karafyllis2009global}, that was also used in \cite{nesic2009explicit}, for which the plant is given by $\dot{x}_P =  dx_P^2-x_P^3 + u$ for some unknown parameter $d$ with $\abs{d} \leq 1$. The controller is given by $u = -2\hat{x}_P$, where $\hat{x}_P$ is the most recently received value of $x_P$. Note that, even though our plant model  \eqref{eq_plant}, \eqref{eq_controller} is not stated in such a generality, this setup can easily be modeled as an NCS of the form~\eqref{eq_NCS1}-\eqref{eq_NCS3} as it has been demonstrated in \cite{nesic2009explicit}.

For $e \defeq \hat{x}_P-x_P$ and $x \defeq x_P$, we obtain $f(x,e) = -2x+dx^2-x^3-2e$ and $g(x,e) = -f(x,e)$. We chose $W(e) \defeq \abs{e}$, which satisfies for all $e \neq 0$ and any fixed \change{}{\changet{parameter}{}} $k\in\left[0,2\right)$
\begin{align*}
	\left\langle \frac{\partial W(e)}{\partial e},g(x,e) \right\rangle&= \text{sign}(e) g(x,e) \\ &\leq \abs{2x+dx^2-x^3-2e}\\ &\leq \change{L(k)}{L_k} W(e) + \change{H(x,e,k)}{H_k(x,e)}
\end{align*}
where $\change{H(x,e,k)}{H_k(x,e)} = \abs{2x+dx^2-x^3-ke}$ and $\change{L(k)}{L_k} = 2-k$. \changet{}{Here $k$ is a tunable parameter that can be varied to obtain a large MATI bound.} \changet{Thus}{As a result}, \eqref{eq_w_bound} and \eqref{eq_h} hold globally. Since the system has only one node,  \eqref{eq_w_jump} even holds  with $\lambda = 0$ and therefore we employ the MATI bound for sampled-data systems from \eqref{eq_bound_sd}. It hence remains, depending on \change{}{the tunable parameter } $k$, to find a preferably small value $\change{\gamma(k)}{\gamma_k}$, such that \eqref{eq_V_bound} is satisfied. We chose  $\varrho(s) = \delta^2 s^2$ for some fixed $\delta >0$. Note that $\delta^2$ determines the worst-case convergence speed for the closed-loop system. We can rewrite \eqref{eq_V_bound} \change{}{for polynomial $V(x)$ }as\change{}{\begin{align}
	\nonumber
	-\left\langle \nabla V(x),f(x,e)\right\rangle - \delta^2 x^2 - \delta^2 e^2&\\
\nonumber	-\left(2x+dx^2-x^3-ke\right)^2 + \change{\gamma^2(k)}{\gamma_k^2}e^2&\\
		= p_1(x,e)d^2+p_2(x,e)d+p_3(x,e) &\geq 0, \label{eq_poly}
	\end{align} where  $p_1(x,e), p_2(x,e)$ and $p_3(x,e)$ are polynomials in $x$ and $e$. 
Therefore \eqref{eq_poly} holds for polynomial $V(x)$, if its left hand side is sum of squares (SOS).  }
We can \change{thus}{moreover } conclude that the left-hand side of \eqref{eq_poly} is SOS for any $d$ with $\abs{d} \leq 1$, if \change{the right-hand side of \eqref{eq_d_poly}}{it} is SOS for all the combinations from $\left(d,d^2\right) \in \left\lbrace \left(1,0\right),\left(1,1\right),\left(-1,0\right),\left(-1,1\right)\right\rbrace$. Note that this \change{approach}{procedure } is inspired by the second example from \cite{omran2016stability}. 

Since we consider a sampled-data system, we will now compare our approach with the approach from \cite{nesic2009explicit}, which is the sampled-data version of \cite{carnevale2007lyapunov}. As in \cite{nesic2009explicit}, we consider $\delta = 0.1$. 
Using Theorem~\ref{theo_stab}, \changet{stability}{UGAS} can be guaranteed for any  $\mati < 0.7909$, which results from $\gamma = 1.544$ and $L = 0.738$, for which \eqref{eq_poly} and thus also \eqref{eq_V_bound} can be verified using SOSTOOLS and the above described procedure with $V(x) = 0.3578 x^4 +1.431 x^2$.
In \cite{nesic2009explicit}, the bound on the MATI was given for the considered example as $\mati \leq 0.368 \si{s}$. To determine $\gamma$ and $L$, an approach that includes some conservative estimates was used in \cite{nesic2009explicit}. These estimates can be circumvented by the above described procedure based on SOSTOOLS. Therefore the best value for the bound on $\mati$ for the approach from \cite{nesic2009explicit}, that we could find %
is $\mati < 0.4762 \si{s}$, which is attained for $\gamma = 2.151$ and $L = 2$ with $V(x) = 0.5 x^4+2 x^2$. Thus, our approach leads to an improvement of more than 66~\percent in comparison to the best value that we could find for the approach from   \cite{nesic2009explicit}. A second example for NCS with multiple nodes and an even higher improvement of the MATI will be given in the next section.

\section{The Linear Case}
\label{sec_lin}
For general nonlinear systems, it is a challenging task to find the smallest possible values for $\gamma$ and $L$ for which Assumption~\ref{ass_hybrid} holds. For the special case of linear systems, \change{}{of the form\change{}{\footnote{\change{}{\changet{Details on $A,C,E$ and $F$  for linear plants can, e.g., be found in \cite{walsh2002stability}.}{For details on $A,C,E$ and $F$ see \cite{walsh2002stability}.}}}}
	\begin{equation}
	\label{eq_sys_lin}
	f(x,e) = Ax+Ee,~g(x,e) = Cx+Fe.
	\end{equation}
	} we can however state a systematic procedure for finding $\gamma$ and $L$ based on an LMI. The procedure is inspired by \cite{heijmans2017computing}. 
\change{We consider in this section linear NCS of the form}{}
\subsection{A systematic procedure to compute $\gamma$ and $L$}
\label{sub_lin_prod}
We shall reformulate \eqref{eq_w_bound} and \eqref{eq_V_bound} in terms of an LMI, that \changet{}{allows} an efficient search for small values of $\gamma$ and $L$ that satisfy Assumption~\ref{ass_hybrid}. The following derivations are inspired by \cite{heijmans2017computing}. We assume that 
	$\norm{\frac{\partial W(\kappa,e)}{\partial e}} \leq M_w$
 and that  $\underline{\alpha}_W(s) = \underline{\tilde{\alpha}}_W s$  for constants $M_w > 0$, $\underline{\tilde{\alpha}}_W > 0$. This is a reasonable assumption for many scheduling protocols, \change{}{cf. \cite[Section V]{heemels2010networked}}. For RR and TOD, it holds with $M_{w,RR} = \sqrt{l}$, $M_{w,TOD} = 1$ and  $\underline{\tilde{\alpha}}_W = 1$,  cf. \cite{heemels2010networked}.  From \eqref{eq_sys_lin}, we obtain
	$\norm{\dot{e}} = \norm{Cx+Fe} \leq \norm{Cx+kFe} + \norm{(1-k)Fe},$
for any fixed \change{}{parameter } $k\in\left[0,1\right)$, leading to
\begin{equation}
	\label{eq_L_lin}
	\change{L(k)}{L_k} = M_w \underline{\tilde{\alpha}}_W^{-1}\norm{(1-k)F}
\end{equation}   and $\change{H(x,e,k)}{H_k(x,e)} = M_w\norm{Cx+kFe}$ in \eqref{eq_h}. We consider $\varrho(s) = \delta^2 s^2$ for some chosen $\delta > 0$. Note that $\delta^2$ determines again the worst-case convergence speed for the closed-loop system. We assume for simplicity\footnote{Otherwise, the NCS would be \changet{uniformly globally asymptotically stable}{UGES} for any transmission interval.} that $\change{\gamma^2(k)}{\gamma_k^2} \geq \delta^2$, implying
	$\left(\change{\gamma^2(k)}{\gamma_k^2}-\delta^2\right) W\change{}{^2}(e) \geq  \left(\change{\gamma^2(k)}{\gamma_k^2}-\delta^2\right) \underline{\tilde{\alpha}}_W^2 \norm{e}^2$
and thus we obtain for the right-hand side of \eqref{eq_V_bound}
\begin{align*}
&-\varrho(\norm{x}) - \varrho(W(\kappa,e)) - \change{H^2(x,e,k)}{H_k^2(x,e)} + \change{\gamma^2(k)}{\gamma_k^2} W^2(\kappa,e)\\
	 \geq&-\delta^2\norm{x}^2 + \underline{\tilde{\alpha}}_W^2\left(\change{\gamma^2(k)}{\gamma_k^2}-\delta^2\right) \norm{e}^2 - \change{H^2(x,e,k)}{H_k^2(x,e)}\\
	 =&\begin{bmatrix}
	x\\e
	\end{bmatrix}^\top J(k) \begin{bmatrix}
	x\\e
	\end{bmatrix}
\end{align*}
 with
\begin{align*}
&J(k)\defeq\\ &\begin{bmatrix}
-\delta^2 I_{n_x}-M_w^2C^\top C & -kM_w^2C^\top F\\-kM_w^2F^\top C& \underline{\tilde{\alpha}}_W^2(\change{\gamma^2(k)}{\gamma_k^2}-\delta^2) I_{n_e}- M_w^2k^2F^\top F
\end{bmatrix}.
\end{align*}
Moreover, for $V(x) = x^\top P x$ with $P\in \mathbb{R}^{n_x\times n_x}$, we obtain for the left-hand side of \eqref{eq_V_bound}
	$\left\langle \nabla V(x), Ax+Ee \right\rangle = x^\top\left(A^\top P + PA\right)x+2x^\top P E e.$
Therefore, \eqref{eq_V_bound} holds for the linear setup \eqref{eq_sys_lin} if
\begin{equation}
\label{eq_lmi}
	\begin{bmatrix}
		\Lambda_{11} & \Lambda_{21}\\
		\Lambda_{12} & \Lambda_{22},
	\end{bmatrix} \leq 0,
\end{equation}
where $\Lambda_{11} = 	A^\top P + PA +\delta^2 I_{n_x}+M_w^2C^\top C,~ \Lambda_{21}=kM_w^2C^\top F+PE,~ \Lambda_{12}=\Lambda_{21}^\top$ and $\Lambda_{22}=-\underline{\tilde{\alpha}}_W^2(\change{\gamma^2(k)}{\gamma_k^2}-\delta^2) I_{n_e}+M_w^2k^2F^\top F$.
\changet{}{This leads to the following corollary of Theorem~\ref{theo_stab}.
	\begin{coro}
		\label{coro_lin}
		Assume \eqref{eq_w_bound} and \eqref{eq_w_jump} hold for some $W, \underline{\alpha}_w$ and $\overline{\alpha}_w$ with $\norm{\frac{\partial W(\kappa,e)}{\partial e}} \leq M_w$, $\underline{\alpha}_W(s) = \underline{\tilde{\alpha}}_W s$ and $\overline{\alpha}_W(s) = \overline{\tilde{\alpha}}_W s$ for constants $M_w > 0$, $\underline{\tilde{\alpha}}_W > 0$ and $\overline{\tilde{\alpha}}_W > 0$.
		Suppose $f$ and $g$ satisfy \eqref{eq_sys_lin} and let \eqref{eq_lmi} hold for some $\gamma_k > 0$, $P \succ 0$, $\delta > 0$ and $k \in \left[0,1\right)$.
		If $\mati$ in \eqref{eq_hybrid_sys} satisfies for $L = L_k$ according to \eqref{eq_L_lin} and $\gamma = \gamma_k$ the bound \eqref{eq_bound} and $0< \epsilon \leq \mati$ then, for the system \eqref{eq_hybrid_sys}, the set $\left\lbrace \left(x,e,\tau,\kappa\right): x = 0, e= 0 \right\rbrace$ is UGES. 
	\end{coro}
	\begin{pf}
		Follows with the preceding derivations from Theorem~\ref{theo_stab}. \hfill\hfill\qed
	\end{pf} 
	}

For \change{}{any } fixed \change{}{parameter } $k$, we can \changet{thus}{}find a suitable value for $\change{\gamma(k)}{\gamma_k}$  %
by minimizing it subject to the LMI constraint \eqref{eq_lmi}. $P$ can be used as an additional decision variable. To find preferably \textit{good} values for $\change{\gamma(k)}{\gamma_k}$ and $\change{L(k)}{L_k}$\changet{ that satisfy Assumption~\ref{ass_hybrid},}{,} we can therefore test different values $k\in\left[0,1\right)$ and compute $\change{L(k)}{L_k}$ and $\change{\gamma(k)}{\gamma_k}$ for each of these values. To select the values of $k$ for testing, for example a uniform grid can be used. For each pair $\left(\change{\gamma(k)}{\gamma_k}, \change{L(k)}{L_k}\right)$, a bound on $\mati$ can then be computed with \eqref{eq_bound}, and we can therefore use the pair $\left(\change{\gamma(k)}{\gamma_k}, \change{L(k)}{L_k}\right)$ that leads to the largest bound. \change{For $k=0$, \eqref{eq_lmi} reduces to the LMI that was considered in \cite[Theorem~3]{heijmans2017computing}, which can be used to determine the MATI 
	bound for the approach from \cite{carnevale2007lyapunov}.}{\begin{rema}
		For $k=0$, \eqref{eq_lmi} reduces to the LMI that was considered in \cite[Theorem~3]{heijmans2017computing}, which can be used to determine the MATI 
		bound for the approach from \cite{carnevale2007lyapunov}. Computing $\gamma_k$ and $L_k$ for any fixed parameter $k$ requires hence solving an LMI of the same dimension as for the approach from \cite{carnevale2007lyapunov}. Thus the computational effort for our proposed approach scales, in comparison to the approach from \cite{carnevale2007lyapunov}, linearly with the number of grid points that are considered.
	\end{rema}
	} 

\subsection{Example\change{}{~2}}
To illustrate our approach for linear systems, we consider the numerical example from \cite{heijmans2017computing}, for which plant and controller are given by
$\dot{x}_P = A_P x_P +B_P u$ and $u = -K\hat{x}_P$ with $A_P = \frac{1}{5} \begin{pmatrix}
-4 & 1\\ -2 & 3
\end{pmatrix},
B_P = \begin{pmatrix}
-1\\2
\end{pmatrix},
$
 and $K = \begin{pmatrix}
 -0.2 & 0.5
 \end{pmatrix}$, where $\hat{x}_P$ contains the most recently received values for the plant states.
 Each plant state is measured by a separate node. For $e = \hat{x}_P-x_P$, we obtain therefore  $A = A_P-B_PK$, $E = -B_PK, C = -A$, $F = -E$ and $l = 2$. Moreover, when we consider the TOD protocol, we can use $\lambda = \sqrt{\frac{l-1}{l}}$ and \changet{}{$W(\kappa,e) = \norm{e}$ with} $\underline{\tilde{\alpha}}_W = M_w = 1$ (cf. \cite{heemels2010networked}). 
 
A comparison of our approach with the approach from \cite{carnevale2007lyapunov} is given in Table~\ref{tab_lin}. In the first column of Table~\ref{tab_lin}, the bound on $\mati$ \change{from}{for the approach from } \cite{carnevale2007lyapunov} is given (\change{}{i.e., the value for $k = 0$, } cf. also \cite{heijmans2017computing}). In the second column of Table~\ref{tab_lin}, we state values for the bound on $\mati$ according to \changet{Theorem~\ref{theo_stab}}{Corollary~\ref{coro_lin}}. To find these bounds on the MATI, we have gridded the interval $k\in\left[0,1\right)$ uniformly with a step size of $0.001$ and minimized $\change{\gamma(k)}{\gamma_k}$ subject to the constraint \eqref{eq_lmi} for each value of $k$ from this grid, using YALMIP~\cite{lofberg2004yalmip}  and Sedumi~1.3~\cite{sturm1999sedumi}.
 Then, we have computed the bound on $\mati$ from \eqref{eq_bound} for each resulting combination of $\change{\gamma(k)}{\gamma_k}$ and $\change{L(k)}{L_k}$. 
 The value for the bound on $\mati$ in Table~\ref{tab_lin} is for each $\delta$ the largest bound that we found using this procedure.  In the last column of Table~\ref{tab_lin}, the improvement for the bound on $\mati$, that can be achieved using our approach, is given. 
 It can be seen that an improvement of \change{more than}{over} 100\percent~is possible in some cases.  %
 \begin{table*}[h]
 	\caption{Comparison of our bound on $\mati$ to the bound from \cite{carnevale2007lyapunov}.}
 	\label{tab_lin}
 	\centering
 	\begin{tabular}{| c |c |c|c|c|  }
 		\hline
		$\delta$ & $\mati$, \cite{carnevale2007lyapunov}  &$\mati$, Corollary~\ref{coro_lin} & $k$ & Improvement \\ \hline
 		2 & 0.044 & 0.0536 &0.999 &20\percent \\ 
 		1 & 0.0743 & 0.1071&0.999 &44\percent \\
 		0.5& 0.1071 & 0.2141&0.999 &99\percent\\
 		0.2 & 0.1337 & 0.2785&0.916 &108\percent\\
 		0.1 & 0.1399 & 0.2817&0.983&101\percent\\\hline
 	\end{tabular}
 \end{table*}
\section{Conclusion}
\label{sec_conc}
We have proposed a simple, yet efficient modification for the approach to compute the MATI from \cite{carnevale2007lyapunov,nesic2009explicit}. The key feature is a slightly more general version of the main assumption from \cite{carnevale2007lyapunov}. Due to this modification, the approach that we have presented in this paper can result in significantly larger bounds on the MATI, as it was demonstrated with two examples from the literature. An improvement of more than 100\percent~is possible in some cases. \change{It seems likely that the proposed modification can also be applied to the setups of many works that are based on the results of \cite{carnevale2007lyapunov,nesic2009explicit}, providing similar improvements there. }{This modification can also easily be applied to the setups of many works that are based on the results of \cite{carnevale2007lyapunov,nesic2009explicit}, providing there similar improvements. }
Moreover, combining our results with different approaches to improve the  MATI bound as, e.g., those from \cite{hertneck20stability,heijmans2020average,heijmans2017computing} \change{may}{can } lead to an even further improvement of the MATI bound.
	\bibliography{../../../../Literatur/literature3}
	\normalsize
	\appendix
\section{Proof of Theorem~1}
\label{append_a}
\begin{pf}
\change{This proof follows the same lines as the proof of Theorem~1 from \cite{carnevale2007lyapunov}, and requires only minor modifications. }{ 
This proof is essentially the same as the proof of \cite[Theorem~1]{carnevale2007lyapunov}, and requires only minor modifications. } Nevertheless, we state it here for the sake of completeness. We consider the solution $\phi:\left[0,\mati\right] \rightarrow \mathbb{R}$ to
\begin{equation}
	 \label{eq_phi}
	\dot{\phi} = -2L\phi - \gamma\left(\phi^2+1\right),\quad \phi(0) = \lambda^{-1}.
\end{equation}
From \cite{carnevale2007lyapunov}, we know that $\phi(\tau) \in \left[\lambda,\lambda^{-1}\right]$ for all $\tau \in \left[0, \mati \right]$. We denote $\xi \defeq \left[x^\top,e^\top,\tau,\kappa\right]^\top$ and $F(\xi)  \defeq \left[f(x,e)^\top,g(x,e)^\top,1,0\right]^\top$, and use subsequently the function 
	$U(\xi) \defeq V(x)+ \gamma \phi(\tau)W^2(\kappa,e).$
Note that 
\begin{align}
\label{eq_u_jump}
	\nonumber U(\xi^+) &= V(x^+) + \gamma \phi(\tau^+) W^2(\kappa^+,e^+)\\
	\nonumber &= V(x) + \gamma \phi(0) W^2(\kappa+1,h(\kappa,e))\\
	&\leq V(x) + \gamma\lambda W^2(\kappa,e) \leq U(\xi).
\end{align}

We also have\footnote{ As in \cite{carnevale2007lyapunov}, we use here  $\left\langle \nabla U(\xi),F(\xi) \right\rangle$ by a slight abuse of notation, even though $W$ is not differentiable with respect to $\kappa$. This is  justified since the corresponding component of $F(\xi)$ is zero.} using \eqref{eq_h}, \eqref{eq_V_bound}  and \eqref{eq_phi}, for all $(\tau,\kappa)$ and almost all $(x,e)$ that
\begin{align}
	\nonumber \left\langle \nabla U(\xi),F(\xi) \right\rangle \leq &- \varrho\left(\norm{x}\right) - \varrho(W(\kappa,e)) - H^2(x,e)\\ \ \nonumber &+ \gamma^2 W^2(\kappa,e) \\
	\nonumber&+ 2\gamma\phi(\tau) W(\kappa,e) \left(LW(\kappa,e) + H(x,e)\right)\\
	 \nonumber &-\gamma W^2(\kappa,e)\left(2L\phi(\tau) + \gamma \left(\phi^2(\tau)+1\right)\right)\\
	 \nonumber \leq &-\varrho(\norm{x}) - \varrho(W(\kappa,e)) - H^2(x,e) \\
	 \nonumber &+2\gamma\phi(\tau)W(\kappa,e) H(x,e)\\
	 \nonumber &- \gamma^2W^2(\kappa,e) \phi^2(\tau)\\
	 \leq &-\varrho(\norm{x}) - \varrho (W(\kappa,e)). \label{eq_same}
\end{align}
Note that \change{this}{\eqref{eq_same} } is the only part of this proof where $H$ and therefore the modified part of Assumption~\ref{ass_hybrid} come into play \change{}{and thus the only part of this proof that is different to the proof of \cite[Theorem~1]{carnevale2007lyapunov}}. The function $\varrho$ is positive definite, and $V$ is positive definite and radially unbounded. Therefore, since $\phi(\tau) \in \left[\lambda, \lambda^{-1}\right]$ for $\tau\in\left[\epsilon,\mati\right]$, there exists a continuous, positive definite function $\tilde{\varrho}$ such that 
	$\left\langle \nabla U(\xi),F(\xi) \right\rangle \leq - \tilde{\varrho} (U(\xi))$.
This implies, using standard arguments for continuous-time systems (see e.g. \cite[p. 146]{khalil2002nonlinear}), that there is $\beta \in \mathcal{K}\mathcal{L}$ such that 
\begin{equation}
\label{eq_u_KL}
	U(\xi(t,j)) \leq \beta(U(\xi(t_j,j)),t-t_j), \forall (t_j,j) \preceq (t,j) \in \text{dom}~\xi,
\end{equation}
where $(t_j,j) \preceq (t,j)$ means $t_j \leq t$, with $\beta$ satisfying also
\begin{align}
\nonumber
	\beta(s,t_1+t_2) = \beta(\beta(s,t_1),t_2),&\\ \forall(s,t_1,t_2)& \in \mathbb{R}_{\geq 0} \times \mathbb{R}_{\geq 0} \times \mathbb{R}_{\geq 0}.\label{eq_beta}
\end{align}
We observe from \eqref{eq_u_jump} that
	$U(\xi(t_{j+1},j+1)) \leq U(\xi(t_{j+1},j))$
for all $j$ such that $(t,j) \in \text{dom}~\xi$ for some $t \geq 0$. Using in addition \eqref{eq_u_KL} and \eqref{eq_beta}, we obtain
	$U(\xi(t,j)) \leq \beta(U(\xi(0,0)),t),~\forall (t,j) \in \text{dom}~\xi,$ 
and thus, as $t_{j+1} - t_j\geq\epsilon$,
	$U(\xi(t,j)) \leq \beta(U(\xi(0,0)),0.5t+0.5\epsilon j),~\forall (t,j) \in \text{dom}(\xi).$
Now, since $V$ is positive definite, using additionally the bounds on $W(e)$ from \eqref{eq_w_bound} and the fact that $\phi(\tau) \in \left[\lambda, \lambda^{-1}\right]$ for $\tau\in\left[0,\mati\right]$, global uniform asymptotic stability of the set $\left\lbrace (x,e,\tau,\kappa): x = 0, e = 0 \right\rbrace$ follows. 

To proof  uniform exponential stability, the additional assumptions for the theorem can be used to show that $\tilde{\varrho}$ can be chosen linear, which allows us  to chose $\beta(s,t) = Ms\exp{-\lambda t}$ in \eqref{eq_u_KL} for some $\lambda>0$ and some $M >0$. This guarantees together with the additional quadratic bounds on $V(x)$ and $W(e)$,  uniform exponential stability. \hfill\hfill \qed
\end{pf}
	
\end{document}